\documentclass{sig-alternate}

\usepackage{amsmath}
\usepackage{amssymb}
\usepackage{color}
\usepackage{dcolumn}
\usepackage{float}
\usepackage{graphicx}
\usepackage[latin9]{inputenc}
\usepackage{multirow}
\usepackage{rotating}
\usepackage{subfigure}
\usepackage{psfrag}
\usepackage{tabularx}
\usepackage[hyphens]{url}

\usepackage[bookmarks, bookmarksopen, bookmarksnumbered]{hyperref}
\usepackage[all]{hypcap}
\newcommand{\code}[1]{\texttt{#1}}

\sloppypar

\newfloat{example}{tbp}{ex}
\floatname{example}{Example}

\begin{document}

\title{Ensuring Correctness at the Application Level: A Software Framework Approach}
\numberofauthors{4}

\author{
%
%
\alignauthor
Eloisa Bentivegna\\
       \affaddr{Center for Computation \& Technology}\\
       \affaddr{Louisiana State University}\\
       \affaddr{Baton Rouge, LA, USA}\\
       \email{bentivegna@cct.lsu.edu}
\alignauthor
Gabrielle Allen\\
       \affaddr{Center for Computation \& Technology}\\
       \affaddr{Department of Computer Science}\\
       \affaddr{Louisiana State University}\\
       \affaddr{Baton Rouge, LA, USA}\\
       \email{gallen@cct.lsu.edu}
\and
\alignauthor
Oleg Korobkin\\
       \affaddr{Center for Computation \& Technology}\\
       \affaddr{Department of Physics \& Astronomy}\\
       \affaddr{Louisiana State University}\\
       \affaddr{Baton Rouge, LA, USA}\\
       \email{korobkin@cct.lsu.edu}
\alignauthor
Erik Schnetter\\
       \affaddr{Center for Computation \& Technology}\\
       \affaddr{Department of Physics \& Astronomy}\\
       \affaddr{Louisiana State University}\\
       \affaddr{Baton Rouge, LA, USA}\\
       \email{schnetter@cct.lsu.edu}
}

\maketitle

\begin{abstract}
As scientific applications extend to the simulation of 
more and more complex systems, they involve an 
increasing number of abstraction levels, at each of which
errors can emerge and across which they can propagate;
tools for correctness evaluation and enforcement at
every level (from the code level to the application
level) are therefore necessary.
Whilst code-level debugging tools are already a 
well established standard, application-level tools 
are lagging behind, possibly due to their stronger
dependence on the application's details.
In this paper, we describe the programming model 
introduced by the Cactus framework, review the High Performance
Computing (HPC) challenges that Cactus is designed to
address, and illustrate the correctness strategies
that are currently available in Cactus at the 
code, component, and application level. 
\end{abstract}


\section{Introduction}
\label{sec:intro}

Modern scientific simulation codes are increasingly complex, 
involving many software components that are combined together 
through workflow tools or frameworks
to investigate multi-physics and multi-scale 
problems. The nature of these problems necessitates
 complex data structures and coupling mechanisms, and the use of
  leading-edge petascale computational environments
with heterogeneous hardware and distributed grid services. 
These issues present 
real challenges for software verification and validation, and the related need for 
code debugging and testing. 

Achieving accurate physical results from computational simulation is a fundamentally
multi-level task, that extends from ensuring faultless elementary 
operations all the way up to assembling a computerized model
that faithfully mirrors the desired physical processes. 
The amount of source code needed to implement software to  investigate real-world  scientific problems leads to abstraction layers
at different levels, causing challenges for software debugging, where the 
 physical model (often represented in a high-level, objected 
oriented design) is separated from the source of the incorrect behavior,
which may lie deep within array operations or memory handling routines. Further, the lack of connection 
between verification methods targeted at different layers both introduces additional complexity and
misses out on opportunities to provide improved paradigms where high-level knowledge about a simulation
can enhance verification at lower levels and vice versa.


There is already a large body of work in computer science on formal 
methods for software verification and validation (see, for instance, 
\cite{Sargent:1999,Pace:2004}). However this research
is not well connected or applied to challenges in high performance computing and 
scientific computing where parallelization, legacy languages (e.g.\ Fortran), and 
application-level issues are not addressed. Additional issues for correctness are
also present at the application level: for example, a piece of software may technically give
the correct result but be implemented in a manner which makes it impractical to run 
on parallel computers --- a particular algorithm could be termed incorrect if it should be
better optimized.

Ideally, debugging tools should be aware of the application as a
whole.  They need to address different levels of complexity, ranging
from simple syntax errors within routines to errors in components and
in the connection of components at run time.  They also need to
address different domains of errors, such as programming errors, errors
in physics equations, in their discretization, or even in the physical
model (approximation) that is used.  These errors may occur at compile
time or at run time, and may be fatal errors like segmentation faults
or more subtle algorithmic, programming, or scientific modelling issues.
In this sense, there is no clear dividing line
between the programmer and the (scientist) user who runs a simulation.

Not only the nature of errors can be very diverse, but there are also
qualitatively different phases
in the process of ensuring that a computerized model correctly represents
the physical model it is designed to simulate (a process usually referred to
as \emph{verification}).

The software programming paradigm itself should be engineered to 
lead to \emph{error prevention} where possible. For example, software developers
can use simple C assertions or conditionals to  articulate physics-inspired consistency 
checks between the code's input parameters. A sophisticated component framework could 
add such checks automatically based on the component descriptions designed at the application 
level. 

Tools and processes are needed to facilitate \emph{error detection}, for example comparing
the application against known solutions, using convergence testing, or comparing against
 other application codes in the same domain~\cite{Babiuc:2007vr}. 
Regression testing can detect other errors by testing against previous software versions, different
architectures, processor counts or resolutions. 

Once the existence of an error is detected, it can take days, weeks or months to track down 
the actual source of the error, before correcting it. \emph{Error identification} is probably the most important and challenging step 
in the verification process, where we expect to see the biggest benefit from application-level 
tools that are integrated with the lower levels, and have potential not only to 
more quickly and easily identify errors, but also to be able to present back to the developer or user 
information on how to correct the error.
Here, many complementary tools (from simple stack traces to 3D data
visualization, to database committing and querying of simulation metadata) 
can be used.

Corresponding to all these different error features and manifestations,
in this paper we describe the approach taken to integrate multiple 
verification strategies into the Cactus framework \cite{Goodale02a,
  cactusweb1}, a generic component framework
designed for large scale parallel simulation code development.  
Since the Cactus framework is used to solve real-world problems, our
tools need to accompany the programmer and user, beginning from
the design and coding stages up to helping the end user test the
results obtained in simulations. Notice that a high-level integration 
of the debugging tools also facilitates their inclusion in or exclusion
from each given simulation, or even simulation stage; since debugging 
may affect the execution performance (e.g.\ due to additional 
checks and parallel barriers inserted in the code), the possibility
to turn on these tools only when absolutely necessary (for example,
in the middle of a simulation, only after an error has been detected) 
is integral to the efficiency of an HPC framework.

Cactus has been used and developed as a high level programming framework in various 
scientific disciplines since 1997, and already includes a number of capabilities 
for error prevention, detection and identification. Recently, new complexity 
presented by e.g.\ multi-model codes and massively parallel compute resources
motivated an effort to investigate new approaches to application-level 
correctness analysis and tools. 
The Application-Level Performance 
and Correctness Analysis project (Alpaca~\cite{ES-Schnetter2007b, ES-alpacaweb}), 
funded by NSF, focuses on ensuring correctness of 
highly-parallel codes, through the integration
of currently available debugging toolkits with
application-aware tools.
Alpaca is concerned not only with ensuring correctness
of a code's scientific results, but also to ensure that those
results are obtained with optimal methods, through profiling and 
optimization; an analysis similar to the one carried out above and 
advocating multi-level verification tools applies to profiling, 
although we will not discuss it explicitly in this paper.

In Section~\ref{usecase}, a use case is presented to illustrate a 
typical Cactus application in the field of numerical relativity.
In Section~\ref{cactus} we describe the essential system design of 
Cactus, including the specification of components. 
Section~\ref{challenges} provides an overview of the different challenges to
verification posed by high-performance environments. In Section~\ref{strategies} we 
describe in detail the different capabilities currently implemented in Cactus applications that contribute to 
verification and validation, and then Section~\ref{alpaca} details new capabilities that are being 
researched and implemented in current work through the Alpaca project.

\section{Astrophysics Use Case}
\label{usecase}

\begin{figure}[!ht]
\includegraphics[width=0.45\textwidth]{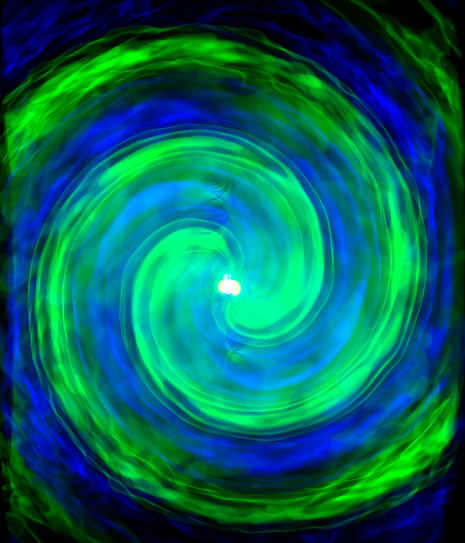}
\caption{Gravitational waves emitted in the coalescence of a pair
of equal-mass black holes. The time-dependent wave amplitude and 
phase are the actual physical observables that relativistic simulations
aim to reproduce, and their validity depends on the correctness of
each single component in the simulation code, from the choice of the 
physical parameters to the evolution of the basic field variables on 
a large number of cores, to the extraction of observables and data 
output (image credit: J.~Ge \& A.~Hutanu).\label{fig:waves}}
\end{figure}

In the field of numerical relativity, more than 15 research groups world-wide have adopted 
Cactus as the underlying parallel framework and community toolkit for their simulation work. One sample group 
is located at the Center for Computation \& Technology at 
Louisiana State University (LSU\@). This team of around ten researchers develops and uses software to  simulate black holes and neutron 
stars. One problem of interest to the group is understanding the 
physics involved in the coalescence of orbiting black holes,
a two-body problem in general relativity.
 In Newtonian gravity, the motion of 
two point particles subject only to their mutual gravitational
attraction has a straightforward solution, but if relativistic
effects are taken into account, the system's dynamics can only
be solved for numerically. Even in the simplest two-body scenario, 
where two black holes interact with each other, emit 
gravitational waves (see Figure~\ref{fig:waves}) and end up merging in a single final 
black hole, the technological requirements are such that only 
recently has the problem become amenable to computational solution.

The black hole code developed by the group implements 
a numerical algorithm for Einstein's equation of general 
relativity, among the most complex equations in physics~\cite{Alcubierre99d, Alcubierre02a}. 
The code is comprised of 60 Cactus modules, which are maintained in several different 
source-code repositories at LSU and at collaborating institutes in Europe. 
The code  involves the evolution in time of 
26 fundamental 3-dimensional variables, with an additional  
41 3-dimensional variables used throughout the code. These variables
are domain decomposed across the computational grid. An additional 
2050 grid arrays or grid scalars are used in the code, primarily in the different
analysis modules, e.g.\ to locate the apparent horizon of the black holes. 

The simulation is run from a parameter file specifying the physical 
and computational details of the run. For the black hole case, around 325
parameters are set in this file, although the number of parameters that could 
be set (default values are used where variables are not set)
is actually much higher.

The code has to deal with accurately resolving features at  
different length and time scales. Moving from the surface of the
black holes themselves to the far field region where 
gravitation radiation is extracted from the computational grid
involves a factor 50 change in length and time scales. 
This requires a mesh refinement strategy 
that can handle this scale range in a dynamical fashion,
where the high resolution regions track the black holes
as they orbit around each other.


The evolution of a binary black hole system, however, does not 
only involve integrating a system of partial differential 
equations along with a mesh refinement algorithm; usually 
a number of diagnostic and postprocessing tools are also acting
during a simulation, analyzing the properties of the black holes,
extracting gravitational waves and calculating their properties, or 
measuring energy
and linear and angular momentum. Corresponding to all these
tasks,
the average Cactus-based simulation of a binary black hole encounter
includes
over two hundreds Fortran, C, or C++ routines between the startup 
and the shutdown phase, and over a hundred routines in 
each iteration of the main loop.

These Cactus-based codes for binary black hole simulations are run 
on a variety of machines worldwide, from simple workstations
to single institution's private clusters, up to the supercomputers 
available through the TeraGrid or other national resources (both in the US and overseas).
Current production simulations harness up to several thousand cores per run, and can take 
weeks to complete using multiple checkpoint-and-recovery phases.

An additional relevant point relates to the origin of the source code. The core framework is supported
by the Cactus team, which employs software engineering principles such as code reviews, regression testing, 
test suites, et cetera. However, the scientists typically use the development version of this software (since these
researchers are typically those driving the development activities of Cactus). Further, the science thorns
used for these simulations are developed either by the local researchers or the general numerical 
relativity community, and involve contributions from graduate students, postdocs as well as software engineers, leading
to a variance in the quality and testing of the software.  


\section{The Cactus framework}
\label{cactus}

Cactus is a component framework designed
for the development of large-scale parallel scientific codes; its
toolkit includes partial differential equation solvers and mesh
refinement packages, along with interfaces to a large number of
third-party libraries including mathematical operations, I/O, and
profiling.
Created in 1997 as a basis for collaborative work in the relativistic
astrophysics community, it has since also found use in a wide range of
other
application areas including astrophysics, quantum gravity, chemical
engineering, Lattice Boltzmann Methods, econometrics, computational
fluid dynamics, and coastal and climate modelling \cite{CS_Ko05a,
  Cactus_Kim04a, Cactus_Rideout:2006zt, Cactus_Major:2005fy,
  CS_Dijkstra_05a, CS_TASC_review, CS_asc_web, CS_Camarda01a}.

Cactus is structured as a
central part, called the \emph{flesh} that provides core routines, and
components, called \emph{thorns}.  The flesh is independent of all
thorns and provides the main program, which parses the parameters
and activates the appropriate thorns, passing control to thorns as
required.  By itself, the flesh does very little science; to do any
computational task the user must compile in thorns and activate them
at run time.

A thorn is the basic working component within Cactus.  All
user-supplied code goes into thorns, which are, by and large,
independent of each other.  Thorns communicate with each other via
calls to the flesh API or, more rarely, custom APIs of other thorns.
The Cactus component model is based upon tightly coupled subroutines
working successively on the same data, although recent changes have
broadened this to allow some element of spatial workflow.  The
connection from a thorn to the flesh or to other thorns is specified
in configuration files that are parsed at compile time and used to
generate glue code that encapsulates the external appearance of a
thorn.  At run time, the executable reads a parameter file that details
which thorns are to be active and specifies values for the control
parameters for these thorns.

User thorns are generally stateless entities; they operate only on
data which are passed to them.  The data flow is managed by the flesh.
This makes for a very robust model where thorns can be tested and
verified independently, and can be combined at run-time in the manner
of a functional programming language.  Furthermore, thorns contain
test cases for unit testing.  Parallelism, communication, load
balancing, memory management, and I/O are handled by a special
component called \emph{driver} which is not part of the flesh and
which can be easily replaced.  The flesh (and the driver) have
complete knowledge about the state of the application, allowing
inspection and introspection through generic APIs.

Cactus applications are designed to execute in parallel on
supercomputer systems.  Efficient parallel execution these days
requires message passing (MPI) \cite{mpiweb}, which is a tedious,
low-level task.  In Cactus, parallelism has been externalized into the
driver, which offers to each thorn only a local view onto part of the
parallel data structures (grid hierarchy) that the driver maintains.  In
doing so, Cactus suggests a certain parallel programming paradigm with
a single, shared, global notion of ``current iteration'' and
``current simulation time''.  This greatly simplifies thorn
programming at the expense of some efficiency.\footnote{There are some
  thorns that use a different paradigm and that use MPI directly.
  These thorns are typically more difficult to understand.}

There are two widely used drivers for Cactus, \code{PUGH} and
\code{Carpet}, both publicly available.  \code{PUGH} provides a highly
efficient parallel implementation of uniform mesh, scaling up to more
than 130k cores.  \code{Carpet} \cite{Schnetter-etal-03b, Schnetter06a,
  carpetweb} offers in addition adaptive mesh refinement (AMR) and
multi-block capabilities that are crucially important for the
astrophysics use case described in Section \ref{usecase} above.

\subsection{Schedule Tree}
\label{sec:scheduling}

One particular notable point in the design of Cactus is the following:
the thorns' configuration files carry sufficient information to permit
\emph{self-assembly} of the thorns that are activated at run time.
That means that it is not necessary to explicitly describe which
thorns are connected to which other thorns; instead, the execution
data flow and execution order are determined by the flesh at run time.
The data flow is determined by giving externally visible names to the
inputs and outputs of thorns (called \emph{variables} for historic
reasons), which corresponds to introducing abstract types for these
inputs and outputs defining possible connections between them.

The execution order is defined by a \emph{schedule} that is determined
only at startup time (a consistency check performed at run time guarantees 
that the specified schedule is in fact valid, and terminates the run if it is not).  
The schedule is derived from each active thorn
specifying certain actions (e.g.\ routine calls) that have to occur at
certain occasions (\emph{schedule points}).  Such occasions could be
e.g.\ performing a time step, applying boundary conditions, or
evaluating a certain quantity that is about to be output.  Thorns can
also specify ordering constraints on the schedule, and can use
\texttt{if} and \texttt{while} statements for conditional or repeated
execution.  The schedule is hierarchical, i.e., thorns can introduce
new schedule points that can be used by other thorns.

\begin{figure} 
  \includegraphics[width=0.5\textwidth, clip=true, trim=20 300 550 10]{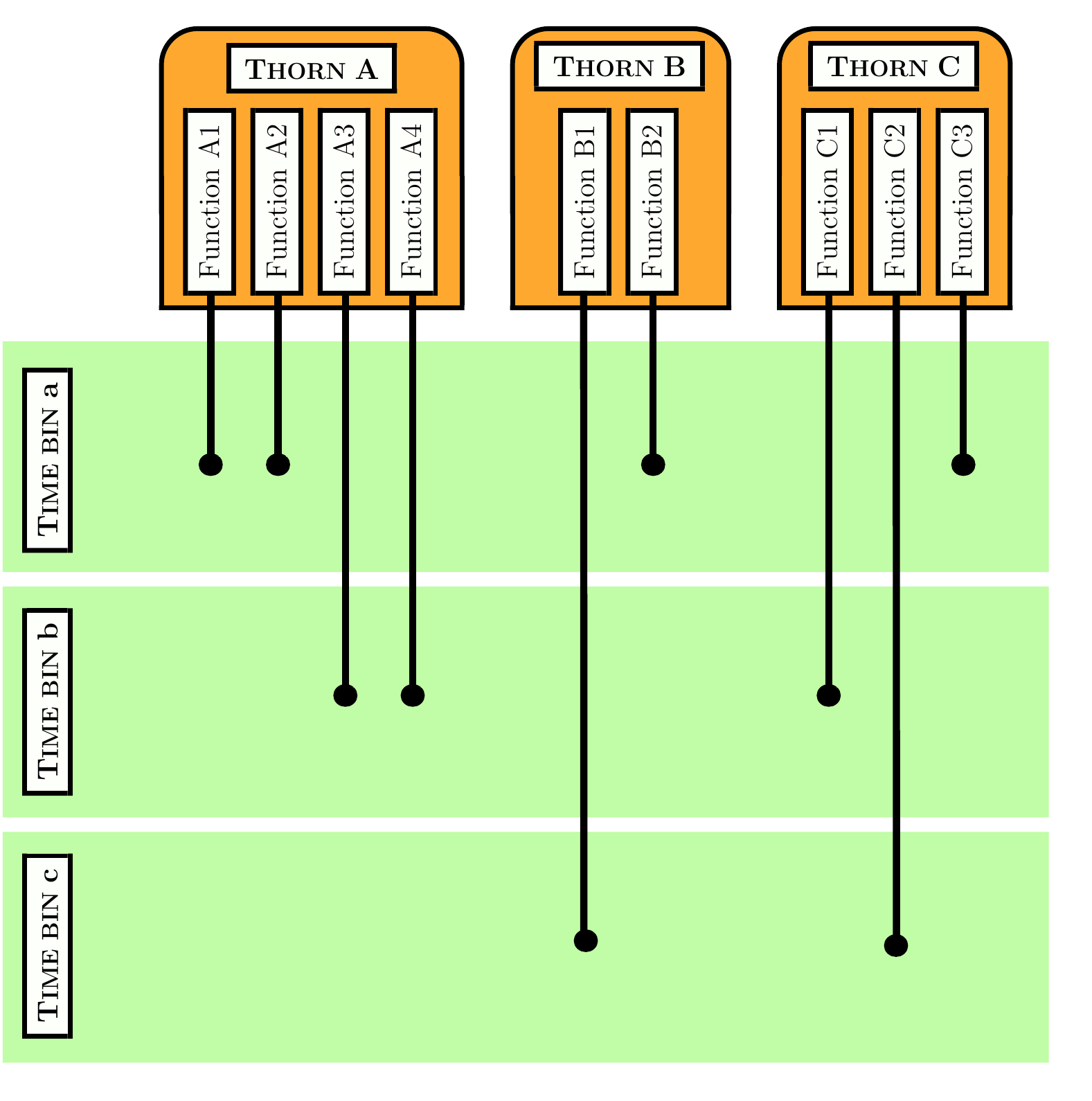}
  \caption{Cactus structures applications from individual thorns which are
    assembled at run time.  The execution order is determined
    automatically, depending on actions and constraints specified by
    each thorn for each schedule point (``time bin'').}
  \label{fig:2D}
\end{figure}

In practice, this means that the schedule is not known until run time,
introducing and added layer of complexity to understanding the
behaviour of the application.  This is exemplified in
Figure~\ref{fig:2D}.

\subsection{Thorn Interfaces, Introspection}
\label{sec:configfiles}

Cactus thorns are described by four configuration files, written in the
Cactus Configuration Language (CCL), that detail the interface exposed
by the thorn.  These are:

\begin{description}
\item[interface.ccl:] Specifies the abstract interface implemented by
  this thorn and lists the other interfaces that this thorn requires.
  It also declares the variables and functions that this thorn uses
  including their types and their accessibility (public or private).
  
\item[schedule.ccl:] Specifies the scheduling directives for the thorn
  (see Section \ref{sec:scheduling} above), describing which routines
  are to be called on what occasion, what variables are used (``need
  storage''), and what variables are modified (``need
  synchronisation'').  Can also contain schedule constraints
  (``before'', ``after''), modifiers (``if'', ``while''), and can
  introduce new schedule points that can be used by other thorns.
  
\item[param.ccl:] Declares run-time parameters for this thorn which
  can be private or public, and declares to which (public) parameters
  from other interfaces this thorn needs access.  This declaration
  includes name, type, allowed ranges, default value, a description,
  and whether it is steerable at run time.
\end{description}

To start a simulation, the user has to provide a \emph{parameter file}
specifying which thorns are to be activated, and giving values for the
run-time parameters.
Only this parameter file determines the set of active thorns and the
execution schedule, and many key characteristics of the application.
A Cactus application is incomplete without such a parameter file, and
it therefore needs to be included in correctness checks.

At run time, the flesh maintains a database containing the thorns'
configurations.  This database is used by the flesh to ensure
consistency both at compile and run time.  In addition, thorns can
introspect the application by querying and/or modifying (if sensible)
the database at run time, allowing thorns to add their own checks.
This is used especially by thorns providing some kind of
infrastructure of their own, e.g.\ the driver thorn mentioned above.

This database allows also to develop generic thorns providing more
targeted debugging capabilities, as described in the following
sections.  This approach is fundamentally different from traditional
debuggers, which insist of inspecting applications from the outside,
since they don't trust the application (after all, it needs to be
debugged!).  In Cactus, one can debug faulty components while still
trusting the flesh to a certain extent.  This makes for a unique
source of high-level information about the application that one can
tap, and which can make it much easier for end users to detect and
locate high-level problems and errors.

For example, a traditional debugger may see independent 3D arrays on
each process, while Cactus ``knows'' that they belong together,
forming a single grid function in an adaptive mesh refinement
hierarchy.  A debugger which trusts the Cactus flesh can allow the
user to examine a 3D visualization of a grid function, or single-step
through the schedule, or modify Cactus parameters in a consistent
manner on all MPI processes at once.

\section{New Challenges for Complex Applications}
\label{challenges}

Modern, large, complex applications pose new challenges to software
development and correctness.  These come in addition to those
challenges that already exist for medium-sized, traditional
applications, and which are addressed by most contemporary software
development environments and debuggers.  These new challenges are in
particular caused by
\begin{itemize}
\item applications that are composed of a large number
  (\textgreater 100) of
  components that have potentially been developed independently;
\item highly parallel execution on large (\textgreater 10k)
  numbers of cores;
\item a large set of run-time parameters, where in effect the
  behaviour of the application is only determined by the end user (and
  not the developer);
\item and, finally, rapidly changing computing environments, where HPC
  systems are considered outdated after only a few years, much sooner
  than the life time of applications.
\end{itemize}

\emph{Large number of components.}  Building a code that simulates a
complex physical system from elementary building blocks inevitably
involves the interaction of many parts, often developed by different
programmers on different platforms, possibly at very different times.
It is important that each component be built with a reusable design in
mind, i.e.\ in a way that makes minimal assumptions about low-level
details such as the underlying environment and the implementation
details of other components of the code, so that both the environment
and the other components can be modified without affecting it.  A
logical separation between a component's internals and its interaction
with the other parts of the code is the base of encapsulation, a
necessary feature of complex codes.  In large teams, encapsulated
components can be developed independently from each other, encouraging
code sharing and reducing duplicated efforts.  Also, as technology,
architectures and algorithms advance, one hopes that it is sufficient
to only improve or replace the outdated components and not the entire
application.

\emph{Large scale parallelization.}  The numerical relativity codes
described in Section~\ref{usecase} currently scale to over 10,000
cores, implementing parallelization by domain decomposition employing
either MPI or hybrid MPI--OpenMP paradigms.  Current funded research
is targeted at achieving scaling to over 100,000 cores necessary to
achieve resolution for the modeling of extremely complex astrophysical 
objects known as Gamma-Ray Bursts.  Designing
software at this level requires complex algorithms for load-balancing
and optimization, and presents challenges for scaling, error
detection, and correction across a diverse set of components.  One
issue presented is simply the vast amount of debugging information
generated on so many cores, and how to present this information in a
meaningful way to developers and end users.  A second issue relates to
the effort and time needed to deploy large scale simulations, which
are expensive in terms of resources used.  The traditional mode of
code debugging is that, after a problem is detected (possibly at late
times into a run), the simulation is run, several times, through a
debugger to interactively investigate and identify problems.  This
approach is not viable when a substantial fraction of a large HPC
system is required for a simulation.  A new paradigm is needed where
useful information is generated while the simulation is running, and
where users can interactively debug running simulations without
needing to rerun.

\emph{Large set of parameters.}  Codes designed to simulate complex
physics have to allow for a correspondingly high number of
user-specified run-time parameters, describing which physical
configuration has to be simulated, which components are required, what
values their options should take, et cetera.  When upwards of hundreds of
parameters are required, a mechanism to check their individual and
combined consistency must be in place.  Depending on the number of
parameters and on the amount of flexibility that the end user is
granted (e.g., can they discretize the spatial domain in a completely
arbitrary fashion, or are they tied to a predefined mesh type?),
consistency conditions may be extremely tricky to formulate.

\emph{Rapidly changing computing environments.}  Programming and
simulation environments available to scientists today are constantly
changing, due to new architectural setups, new message-passing
implementations, new library versions and, from time to time, even new
parallel programming models.  It is therefore important to factor all
low-level details out of the components and organize them in a single
configuration layer that can be adapted to the desired system,
providing the components with an abstract interface to it.

\section{Cactus strategies for correctness}
\label{strategies}

Below we illustrate how various kinds of errors can be detected and
identified in Cactus, both at compile time and at run time.  The
run-time checks include facilities to detect errors at the level of
the source code, at the level of whole components ensuring their
internal consistency, and also at the application level to prevent
potential errors made by the end user actually performing simulations;
as mentioned in the introduction, thanks to steerable parameters many 
of these tools (such as \emph{poisoning}) can be switched on and off while 
the simulation is running. 

Notice that while the vast phenomenology of error sources and 
manifestations demands a similarly articulate debugging approach, with 
distinct strategies acting at the same time, Cactus provides
programming and execution cohesion.

\subsection{Build-Time Mechanisms}
At build time, the Cactus Specification Tool (CST) 
parses all configuration files (CCL files; see Section
\ref{sec:configfiles}) of thorns included in the application and
checks them for consistency, making sure that the required functionalities
are present and that there are no conflicts. 
For instance, the interaction and dependency
of different parts of a Cactus-based code on each other are 
enforced through an \emph{inheritance} mechanism.
Through inheritance, Cactus adds an abstraction layer
providing or denying access to
variables: if thorn $A$ inherits from interface $B$, $A$ will be able to 
access all of $B$'s \emph{public} variables, plus those public 
variables that $B$ may have inherited from other interfaces;
this additional layer allows for a hierarchical
system of information sharing.
The CST also generates the code 
infrastructure to bind the flesh and the thorns together.  These 
automated checks and predefined structure provided by the flesh 
ensure that Cactus components are assembled in the 
correct way, providing a uniform model across collaboration teams
of different composition and background.


A certain level of automation for infrastructure routines
is accomplished by a number of inline functions and
macros that are part of the Cactus flesh and/or
the driver.  These mechanisms offer high-level 
abstractions e.g.\ for passing variables to subroutines, for accessing
grid point data stored in distributed arrays, and for iterating over
the local parts of these arrays; Cactus components can use these as
black boxes.

Another potential source of errors lies in input parameter files. 
As mentioned in Section~\ref{cactus} above, Cactus uses key-value pairs 
for parameters, where each parameter name is qualified by the 
implementation or thorn in which it lives. Thorn authors
specify the parameters of a thorn, which includes the scope, type, 
allowed ranges, default value, and description of each parameter. 
For example, in a thorn setting up astrophysical initial data, 
a parameter \code{central\_density} could be defined in the following way:
\begin{verbatim}
private:
REAL central_density "The star's central density"
{
  (0.0:* :: "The central density must be positive"
} 1.0
\end{verbatim}
which indicates that the parameter \texttt{central\_density} is a real
variable, subject to
the physical constraint of only taking on positive values, with a
value of 1.0
used as default if no value is set in the parameter file.  Further, the parameter 
is private to the thorn and not visible elsewhere.
The standard behaviour of Cactus is to immediately abort with an 
error when the simulation is run if the parameter is outside the 
allowed ranges or does not exist (for example, if the 
parameter name was misspelled in the parameter file).
Cactus can also be run with stricter or more 
relaxed parameter checking setting, for example aborting if a parameter 
is set for a thorn that is not active. 

For long-lasting, expensive simulations that run on large production
machines, errors are not only limited to errors in
writing source code, compiling it, or specifying the correct input
parameters; since designing and carrying out a simulation becomes
such a complex task, mistakes in labelling and handling output files and
in the job submission procedure can also impact the correctness of 
the results.
To address this, the Simulation Factory~\cite{ES-simfactoryweb} 
automates 
many low-level tasks connected with building the executable, setting up, and running a Cactus
simulation, such as maintaining a consistent source tree and handling
compiler options across several machines, selecting the appropriate executable (or creating
a new one) and parameter file, creating new, unique working directories, and submitting
the job for execution, using whichever batch execution system each
specific machine employs. 

Unfortunately, the class of potential errors that are known in advance and 
can be prevented by the strategies outlined above only
begins to cover the sources of problems that plague 
modern simulation codes.
Errors that do not fall in this class must, first and 
foremost, be detected by examining the behaviour of the 
code in a range of cases. This analysis can include
scanning data for non-physical values (infinities, 
NaNs, or out-of-range values such as negative 
temperatures), memory initialization checks such as 
poisoning (see below), asymptotic methods such as Richardson
convergence tests~\cite{Richardson10}, and the traditional 
regression tests. 
Once an error has been detected, it is necessary
(and often laborious) to identify its source. There
are two possible approaches: if the source code for a 
correct version of the code is available, a comparison of files
can highlight the differences and point to the
location of the error; otherwise, the code can be executed step by
step while examining the partial results at end of 
each execution unit, in the hope to isolate the 
origin of the error.

In the following three subsections, we will describe
a number of tools that handle error detection and identification
based on information available at run time at the code level, at the
component level, and at the application level.

\subsection{Code Level Mechanisms}
\begin{figure}
\includegraphics[width=0.5\textwidth]{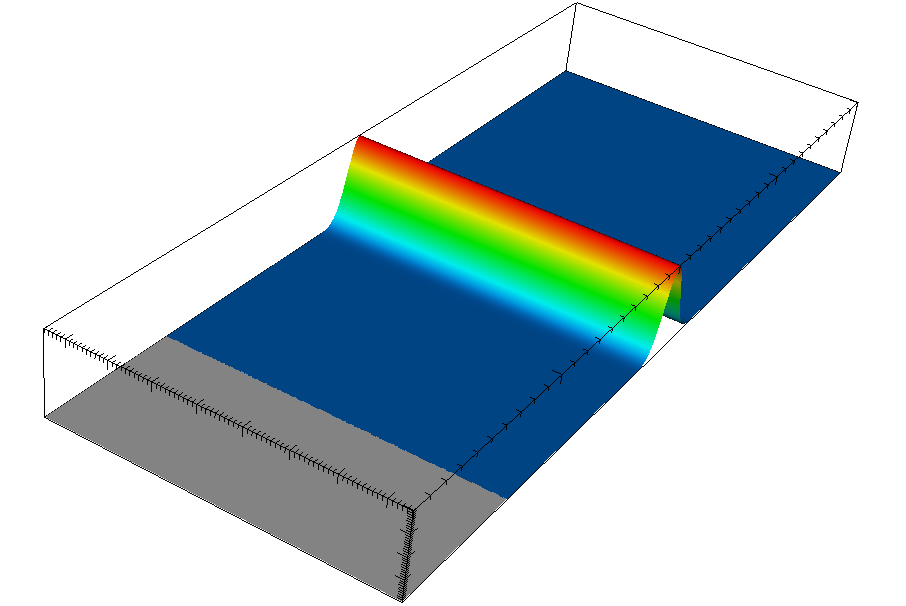}
\caption{
Due to numerical error, a field that is supposed to be 
positive everywhere, like the one shown in the elevation plot above
 may assume small negative values. Inadvertently
taking a square root of this function will generate NaN values.  Their
location can be found by displaying the mask file output by 
thorn \code{NaNChecker} (represented here by the grey shade).\label{fig:nan}}
\end{figure}

One of the first steps towards detecting a simulation 
error is to scan output data for anomalous 
values.  In Cactus the thorn \code{NaNChecker}
will check for infinities and NaNs
at the user-specified intervals.
Through the parameter \code{action\_{\hspace{0em}}if\_{\hspace{0em}}found}, 
the user can prescribe what action is to be performed upon 
detecting either of these values: issuing a simple warning,
terminating the run in a consistent way (e.g.\ after completing
the current iteration), or aborting immediately. The user
can also specify whether a ``mask'' file should be output that describes at
what grid points
in the simulation domain the NaNs were found;
this is often most useful when
investigating the origin of the error.
\begin{description}
\item[Example 1:] A simple code is designed to implement the 
1D advection equation for a positive-definite scalar field $\Phi$ (which could
represent, for instance, the density field in a fluid), and
to compute a derived quantity $F(\Phi) \sim \Phi^{3/2}$ at each
point, which is well defined since $\Phi$ is assumed to be always positive.
However, in the regions where $\Phi$ is comparable to the 
numerical truncation error, it can still assume (small) negative
values, so that the calculation of $F(\Phi)$ yields non-numerical
values. Activating \code{NaNChecker}, the presence of NaN values and their 
locations are announced in the run's standard output, and an HDF5 mask
file is output. An inspection of the NaN locations (Figure~\ref{fig:nan})
reveals that they are concentrated in an area where $\Phi$ is very small, 
hinting that they may be an effect of numerical error.
\end{description}

Once an error has been detected, a traditional method to
debug single routines (especially in
parallel scenarios) is to track the execution by periodically outputting
relevant status information. 
For this purpose, Cactus offers a run-time multi-tier warning system that can be 
paired with the standard assertion checks.  A warning level is 
attached to each condition (with level 0 being the most severe 
level of warning), and the program behavior for each level 
of warning can be customized with a run-time option: executing
a Cactus-generated code with \code{-error-level=N} will turn
all warnings with a level of $N$ or smaller into fatal errors.
In this way, the user can decide which range of 
warnings are harmless and can be tolerated, and which ones 
will definitely compromise a simulation and should therefore
trigger a shutdown.
In addition to the logging data generated
by the Cactus warning system, the verbosity of a large 
class of thorns can be adjusted via run-time parameters, easily
allowing for a detailed analysis of the run's events. Through
a run-time flag, users can also request to receive output from all child 
processes in a parallel environment, in addition to the output from the parent
process.

\subsection{Component Level Mechanisms}
Cactus has a number of tools designed to ensuring the correctness of
whole, individual components. For instance, if a certain version of a
component has been successfully verified to be correct, other
versions, or the same version built on a different system, can be
compared to this version to ensure their correctness.  For this task,
it is crucial that source code is appropriately stored and labelled,
so that the old source version can be retrieved.  Note that version
control systems can often not be used for this task, because they may
only allow write access to a select group of people.  In Cactus, the
thorn \code{Formaline} offers a solution: at compile time, if
\code{Formaline} is in the configuration's thorn list, a set of
\code{tar} balls of the source (one per thorn) are created, and are
added to the executable.  At run time, if thorn \code{Formaline} is
activated, the tarballs are unpacked to recreate the original source
code.

Comparisons with past versions is not only a means to 
trace the origin of errors, but also a way to detect 
unnoticed errors through regression testing, i.e., by comparing
the output produced by two different versions of the code.
In Cactus, regression testing
is automated:
each thorn contains a \code{test} subdirectory with
parameter files and their resulting simulation output, and a single
command starts a regression test of all thorns in a configuration.
Furthermore, we are running nightly build and regression tests for a
large set of
thorns (both infrastructure and physics), displaying test results and
their history on a web portal \cite{ES-portalweb}.

\subsection{Application Level Mechanisms}
In addition to consistency checks related to code structure
and syntax, application-level consistency of data can be 
assessed by a few special thorns, such as the driver or 
the file I/O tools, which are able to verify the integrity 
of data through checksums (for instance, comparing data checksums
before and after an operation gives the user a tight control over
its effect on the checksummed data; this can highlight unwanted 
changes and point to their source).

Application-level information is also available through
the thorn \code{Formaline} described above, which
can, at run time, broadcast execution metadata to an
external server, for real-time monitoring or logging. 
A somewhat complementary approach is provided by the
Simulation Factory, which provides 
a high-level interface for storing simulation metadata
(such as the parameter file, the run host machine, the 
standard output or the data files).

A different approach consists in including, in the 
build process, components that are specifically deputed
to examine that certain conditions hold, or that
certain actions have been performed. The thorn 
\code{CheckSync}, for instance, offers an API that
can be used within other thorns to check that a variable
has been synchronized across parallel processes; the 
thorn \code{CheckTimestepSize} checks that Courant factor
corresponding to the spatial and temporal resolution is 
below a user-specified value for non-uniform grids.

\begin{figure}
\fbox{
\includegraphics[width=0.45\textwidth]{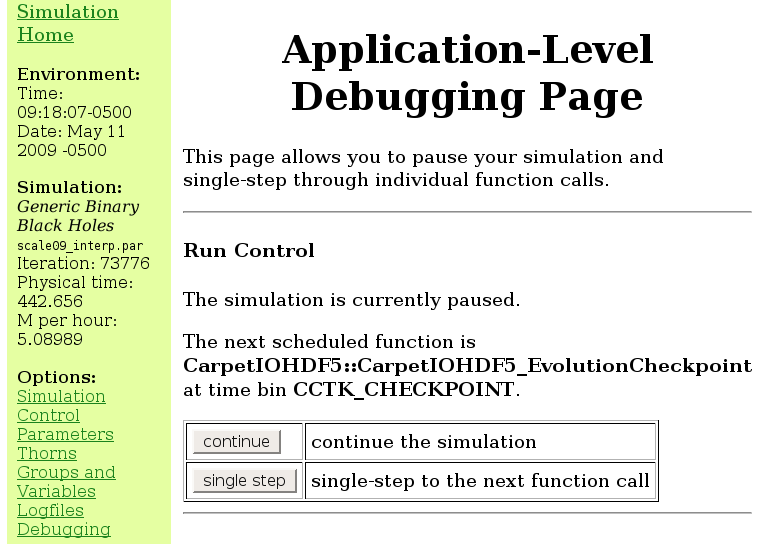}
}
\caption{Web interface to a simulation through the thorn \code{HTTPS};
the figure shows one of the execution control pages.\label{fig:https}}
\end{figure}

When an error is positively identified,
single-step execution of the code
while monitoring the appropriate variables is usually 
an effective method to identify the error's origin. For 
production codes, this strategy
requires tools capable to control the execution and visualize
the data of simulations potentially running on thousands of computing
cores. A Cactus thorn called \code{HTTPS} realizes this functionality
by launching, on the simulation machine, a web server that 
can be interrogated by a remote user and correspondingly steer 
the simulation's execution (see Figure~\ref{fig:https}), provide information on the 
simulation's details, and visualize 1D or 2D slices of its 
output.  This functionality of is currently being extended in the
Alpaca project.  Providing direct, interactive access to applications
executing on large, remote systems is in our opinion one of the key
elements to simplifying the use of HPC resources to solve complex
real-world problems (for the performance of a variant of this tool 
in a real-world scenario, see~\cite{ES-Hutanu2009a}).

Faulty logic and errors in memory management are frequent 
issues in building a large-scale code.  In Cactus, memory management
and parallelism
are assigned to a \emph{driver} thorn, which handles 
allocation and deallocation of variables and manages the 
grid structure. A useful practical tool for tracking 
uninitialized variables (or variables that should have been
re-calculated at a certain point but were not) is \emph{poisoning},
i.e., setting array element to an easily-recognizable (usually
very large) value, so that inspection of the data will
quickly expose those grid points that have not been defined.

\begin{figure}[!t]
  \includegraphics[width=0.5\textwidth, clip=true, trim=150 30 20 80]{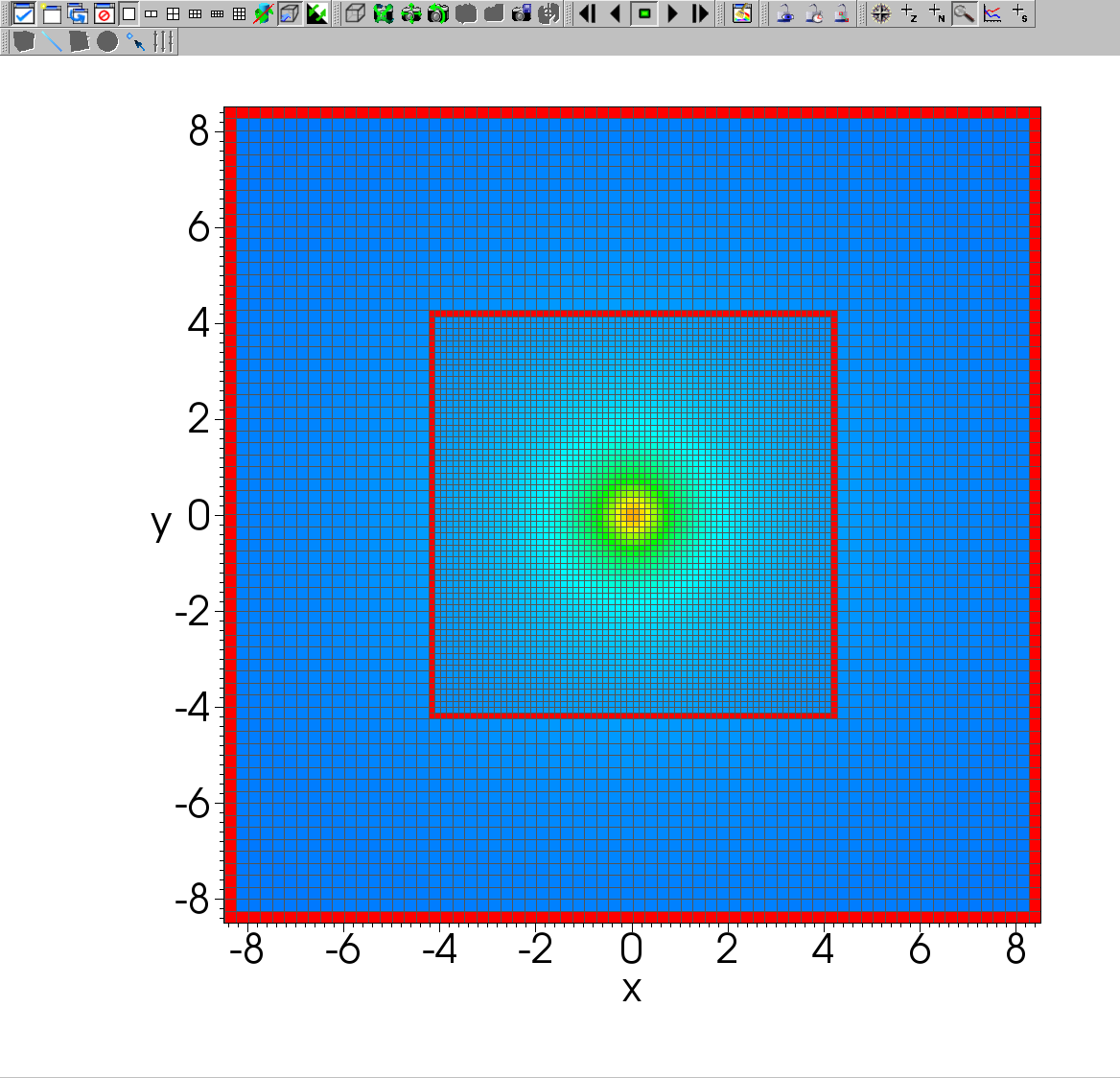}
  \caption{The gravitational field of a single non-rotating black
    hole, on a domain consisting of by two mesh-refinement boxes.  By
    mistake the box boundaries were not initialized, constituting a
    severe error in the execution flow.  By automatically ``poisoning''
    grid functions via Cactus before they are used in the application,
    uninitialized values stand out prominently, here visible as a red
    border around the two boxes (in this case corresponding to a
    nonsensical value of $\sim 2 \cdot 10^6$).}
  \label{fig:poison}
\end{figure}

\begin{description}
\item[Example 2:] In a hydrodynamical simulation of 
an accretion disk on the 
fixed background of a single non-rotating black
hole, the value of the metric tensor (representing the 
gravitational field of the black hole) is set in the 
same loop that updates the dynamical variables 
such as the density and temperature of the disk. The 
loop, however, does not cover the grid boundaries
because the finite differencing stencil cannot be 
employed there. An apposite boundary condition routine 
sets the value of the dynamical variables
on these points, but non-evolving
functions such as the metric tensor are not covered.
With \code{Carpet}'s parameters \code{poison\_new\_memory}
and \code{check\_for\_poison} set to \code{yes}, the user
receives a list of warnings:

\begin{verbatim}
WARNING level 1 ... -> At iteration 0: 
timelevel 1, component 0, map 0, 
refinement level 0 of the variable "gxx" 
contains poison at [0,0,0]}
\end{verbatim}

A visualization of \code{gxx} then shows
the poison locations (Figure~\ref{fig:poison}).
\end{description}

For simulations involving differential equations, Richardson convergence
tests are a useful method to detect and localize errors.
In such tests, 
the dependence of the truncation error (associated 
with the discretization of the differential system) on the resolution
is measured by repeating the simulation with a several different
mesh spacings; one can then compare this value to that expected from the
particular numerical scheme being employed.
\code{Carpet} provides the infrastructure for 
convergence tests, through the \code{convergence\_level},
and \code{convergence\_factor} 
parameters. Adding these two 
parameters to a simulation's
parameter set will automatically modify the simulation's grid spacing,
simplifying the task of managing a group of simulations forming a
convergence test.

Physical correctness can also be verified via other methods,
depending on the equations governing the system. If the system's
evolution is described by a \emph{constrained} set of equations,
monitoring the value of the constraint violation can help the user 
determine whether the numerical solution is an actual physical 
solution for the system.  This is the case for the thorn \code{ADMConstraints}
that computes the value of the constraint violation within the \emph{ADM 
formulation} of general relativity~\cite{Arnowitt62} that we employ.

\section{Conclusions and Future Work}
\label{alpaca}

Although the speed and performance of high-end computers have
increased dramatically over the last decade, the ease of programming
such parallel computers has not progressed.  The time and effort
required to develop and debug scientific software has become the
bottleneck in many areas of science and engineering.  The difficulty
of developing high-performance software is recognized as one of the
most significant challenges today in the effective use of large scale
computers.

This paper has described some of the capabilities already available to Cactus applications
that facilitate verification, debugging and testing of simulations. These tools are widely used by
the Cactus user community, and provide application-level consistency checks, error prevention, 
detection and identification. Some of these tools are automatically invoked by Cactus for any simulation
(e.g.\ parameter checking), other tools can be invoked to investigate particular problems (e.g.\ \code{NaNChecker}).
We are developing additional high-level tools, complementary to
existing debuggers, in the Alpaca project~\cite{ES-Schnetter2007b,
  ES-alpacaweb}.

Software debugging and verification, particularly for large scale scientific applications, is obviously an immensely 
complex task, which needs to be attacked on multiple fronts, from application-level tools as described here, to 
low-level routine debuggers, to new tools to support distributed computing platforms. Despite these envisioned 
improvements, it is still clear that the current complexity of scientific software is providing a barrier to its widespread 
use and development, particularly given that most scientific code development is undertaken by students and postdocs 
working in application-science disciplines. 

One approach that we believe is crucial to address software complexity is to move the programming 
interface to a higher level than that implemented currently in Cactus. Kranc~\cite{kranc04, Husa:2004ip, krancweb} is a tool already used by the LSU relativity 
group to generate entire black hole simulation codes directly from the underlying governing partial differential equations. Kranc uses
Mathematica to generate Cactus thorns from an input file containing the governing equations, fundamental numerical discretizations 
(e.g.\ higher order differencing), and other necessary information
such as boundary conditions and initial data~\cite{ES-Brown2007b, ES-mclachlanweb}; both the source code and the Cactus Configuration
Language files are created. This methodology
makes it straightforward to test new systems of equations and change discretization algorithms, and reduces the introduction of errors. 
Future work for Kranc could involve the automatic integration of techniques for verification, for example tracking the flow of data through a code
or automatically adding metadata information. 

Cactus already contains a good amount of checking of consistency at the computational layer, for example ensuring that 
data types are correct, that methods are called in the specified order, et cetera. However, high-level checking is not routinely present for science 
thorns, for example to check the consistency of evolution schemes with boundary conditions or initial data, or to check for the appropriateness 
of a particular analysis method. Application-level description languages are needed that can encapsulate the entire physical information about 
a simulation, and lead the way for improved automated code generation, reporting,  and data archiving. 

This work does not address validation of data, that is comparison
against physical data~\cite{Sargent:1999,Pace:2004}.
For the black hole problem, only a few exact
solutions to Einstein's equation are known, and are not yet validated
in nature.  Following various past and current initiatives for code
comparisons~\cite{ES-Babiuc2007a, ES-Aylott2009a, ES-Cadonati2009a}),
the next phase will be to verify against observational data from
gravitational wave detectors operating around the world.






\section*{Acknowledgements}

We acknowledge the contributions of the former and current Cactus team in the
design and implementation of the capabilities described in this paper, particularly 
Tom Goodale and Thomas Radke. 
This work is currently supported by NSF awards Alpaca (\#0721915), XiRel (\#0701566) 
and Blue Waters (\#0725070). 
Binary black hole simulations were performed under TeraGrid allocation
TG-MCA02N014.

\bibliographystyle{amsplain-url}
\bibliography{references,vvreferences,publications-schnetter,CCT_CS,Applications}

\end{document}